# Wavefront sensor based on varying transmission filters: theory and expected performance

FRANÇOIS HENAULT*

CRAL - Observatoire de Lyon, 9 Avenue Charles André, 69561 Saint Genis Laval, France

Correspondence *François Hénault. Email: henault@obs.univ-lyon1.fr

Abstract

The use of Wavefront Sensors (WFS) is nowadays fundamental in the field of instrumental optics. This paper discusses the principle of an original and recently proposed new class of WFS. Their principle consists in evaluating the slopes of the wavefront errors by means of varying density filters placed into the image plane of the tested optical system. The device, sometimes called 'optical differentiation WFS' is completed by a digital data-processing system reconstructing the wavefront from the obtained slopes. Various luminous sources of different wavelengths and spectral widths can be employed. The capacities of the method are discussed from the geometrical and Fourier optics points of view, then by means of numerical simulations showing that the ultimate accuracy can be well below λ/10 and λ/100 Peak-to-Valley (PTV) and RMS respectively, provided that certain precautions are taken.

*Keywords:* Wave-front sensing; Fourier optics; Spatial filtering; Phase measurement

## 1 Introduction

The purpose of this paper is to discuss the principle of an original family of Wavefront Sensors (WFS), recently proposed by different authors. The use of such devices is fundamental in most domains of instrumental optics, where they can be employed in extremely different purposes and circumstances, e.g. measurement of single optical components, evaluation of an already integrated optical instrument, and calibration of systems where the wavefront quality evolves with certain physical parameters such as temperature or orientation of the gravity. A few metrology tools such as laser-interferometers or Shack-Hartmann wavefront sensors are nowadays available and commonly used to realize quick and efficient measurements, but their commercial prices remain somewhat high. Wavefront Errors (WFE) may also be rapidly varying with time, for example when propagated from space through turbulent layers of the terrestrial atmosphere and observed at the focus of ground telescopes: this is the astronomical 'seeing', which can be corrected by modern adaptive optics systems requiring specific and fast-frequency WFS. Thus the search for new WFE measurement methods still keeps all its sense.

The basic principle of this alternative class of wavefront sensors consists in evaluating the slopes of the WFE by means of varying density filters directly placed into the image plane of the measured optical system, as will be explained in the next section. The older and simpler of these spatial gradient filters is indeed the Foucault knife-edge [1], allowing the observation of a black and white image of the exit pupil of the tested optics. Because it only requires standard and inexpensive accessories, this technique widely spread in the field of instrumental optics and astronomy and still stays very popular among amateur-astronomers. However the Foucault test is reputed for its limited ability to provide accurate and quantitative data: the produced 'Foucaultgrams' are only readable by experienced operators and only reveal low order aberrations. Following the development of modern computers, Wilson [2] looked for a direct inversion process between the Foucaultgrams and the wavefront errors $\delta(x,y)$ transmitted by the optical system. He demonstrated that such a relationship effectively exists, but is only applicable to wavefront defects of weak amplitude, thus considerably limiting the practical useful domain of the knife-edge test.

Let us mention briefly that due to progresses in electronics and computer technologies, some improvements of the Foucault test, susceptible to provide quantitative data, were proposed by different authors [3-4]. The general idea consisted in determining, for every point M of coordinates (x,y) in the OXY exit pupil plane (see the Figure 1, which are the transverse aberrations $x'(x,y)$ and $y'(x,y)$ of the luminous ray emitted from M in the O'X'Y' image plane. x' and y' are linked to the wavefront error $\delta(x,y)$ by the classical differential relationships [5]:

$$x'(x,y) = D \times \frac{\partial \delta(x,y)}{\partial x}$$
$$y'(x,y) = D \times \frac{\partial \delta(x,y)}{\partial y} \qquad (1)$$

where D is the distance from the exit pupil to the image plane. To determine the transverse aberrations, the knife-edge was displaced in the image plane following the X' and Y' axes until observing the sudden illumination or darkening of the point M(x, y) on the exit pupil. In the field of adaptive optics for astronomy, these revisited versions of the Foucault test recently gave birth to a novel measurement device, the pyramidal wavefront sensor [6], where the displacement of the 'knife-edge' (becoming a four-faces glass pyramid) is replaced with a scanning mirror located in the instrument pupil. But it must be pointed out that most of these developments require mobile equipment and numerous series of pupil image acquisitions and processing. This often makes them time-consuming and difficult to implement, and is probably the reason why they remain rarely used in laboratories or in industry.

To overcome these limitations, different authors proposed to replace the knife-edge with varying transmission filters able to deliver fast, quantitative and accurate results. In 1972, Sprague and Thompson [7] suggested a linear ramp amplitude transmittance filter having the basic property of optically differentiating the objects under observation. Later, Hoffman and Gross [8] then Horwitz [9], proposed different filter shapes, respectively a 'staircase' Foucault knife and a ramp intensity transmission profile. Other filtering functions are obviously possible, but the formula from Sprague and Thompson probably remains the best known and most studied. However the aforementioned authors essentially aimed at improving the contrast of phase-objects in the field of microscopy, and did not seem to realize the potential of their discoveries when applied to the measurement of the WFE of optical systems. The first wavefront sensor explicitly based on a gradient transmission filter was described by Bortz [10] in 1984, and a simplified version of this WFS was studied by Oti [11] in view of its application to adaptive optics. Finally a few experimental studies were conducted by some authors, leading to encouraging results [12-13].

What is the real potential of such 'varying transmission filter' devices for WFE measurements ? Could they really be competitive with modern Shack-Hartmann or laser-interferometers ? This paper attempts to bring some answers. The paragraph 2 shortly summarises the principle of this new class of wavefront sensors, as well as its theoretical basis following the formalism of geometrical and Fourier optics (section 3). The intrinsic accuracy of the method is evaluated by means of several numerical simulations presented in the section 4. Finally, the paragraph 5 gives a brief conclusion about its capacities and future developments.

## 2 Description of the wavefront sensor

The general principle of the wavefront sensor is shown in the Figure 1. It consists in illuminating the optical system to be tested by a pinhole source of light. The source belongs to the object plane that can be placed at finite or infinite distance, respectively producing a spherical or flat reference wavefront at the entrance of the tested optics. The latter forms an image spot of the point source in the O'X'Y' plane, widened and distorted under the influence of the defects and aberrations to be measured. A motionless filter of varying density whose amplitude transmission varies linearly along the transverse X'-axis is then installed near the theoretical image point. The exit pupil of the optical system is imaged on a CCD camera by means of the relay optics represented on the Figure 1. We should then observe a distribution of grey-levels on the exit pupil, whose intensities are directly linked to the transverse aberration x'(x,y) of the tested optical system.

[Figure 1: General principle of the wavefront sensor]

The slopes of the wavefront error along X can be derived by means of a simple relationship: for example, the linear amplitude transmission filter of Sprague and Thompson may be expressed in the image plane as:

$$t_A(x') = \frac{1 + x'/x'_1}{2} \qquad (2)$$

where $x'_1$ is the inverse of the filter slope. Combining equations (1) and (2) and evaluating the observed intensity $I_{Fx}(x,y)$ in the pupil plane easily leads to:

$$I_{Fx}(x,y) = I_N \left[ 1 + \frac{D}{x_1'} \frac{\partial \delta(x,y)}{\partial x} \right]^2 \qquad (3)$$

Here $I_N$ is the uniform intensity measured in the pupil plane when the gradient filter is replaced by a uniform density of 50% amplitude transmission – which should be seen as the photometric calibration procedure of the WFS. The inversion of the previous equation is straightforward and leads to a very simple relationship between the slopes of the wavefront and the observed intensities in the pupil:

$$\frac{\partial \delta(x,y)}{\partial x} = \frac{x_1'}{D} \left\{ \sqrt{\frac{I_{Fx}(x,y)}{I_N}} - 1 \right\} \qquad (4)$$

The same operation can be repeated by turning the filter of 90 degrees around the optical axis, thus allowing the estimation of the slopes along the Y-axis through a similar relationship. The last step of the process finally consists in reconstituting the wavefront error δ(x,y) by double numerical integration from its slopes. Since around thirty years however, due to the developments of the Shack-Hartmann wavefront sensors based on micro-lens arrays, this final operation is well mastered (see for example the paper written by Southwell [14], who described and compared several types of such phase reconstruction algorithms).

An important consequence of the basic relation (4) is that it does not depend on the wavelength λ of the considered light source. Indeed the wavefront error δ(x,y) is here expressed in terms of micrometers, so that its partial derivatives with respect to x and y are homogeneous to angles (expressed in radians), whereas the normalized distribution of intensity $I_{Fx}(x,y)/I_N$ is dimensionless. It can be concluded that if the optical system to be measured is not affected itself with chromatic aberrations, the method can be employed with wide spectrum (or white light) sources, whose energy contribution should allow to obtain high Signal-to-Noise Ratios (SNR) and as a consequence an improved measurement accuracy.

Up to now we followed a simple approach based on elementary considerations of geometrical optics. It is now worthwhile to provide a more rigorous theory of the method, based on the formalism of Fourier optics. This is the scope of the next paragraph.

## 3 Fourier optics analysis

Let us note $B_R(x,y)$ the bidimensional amplitude transmission function in the pupil, uniformly equal to 1 inside a circle of radius R, and zero outside of this circle – this is the 'pillbox' or 'top-hat' function. If λ is the wavelength of the incoming light (assumed to be monochromatic), the wave emerging from the tested optical system can be written at the exit pupil:

$$A_P(x,y) = B_R(x,y) \exp\left[ i \frac{2\pi}{\lambda} \delta(x,y) \right] \tag{5}$$

where δ(x,y) is the wavefront error already defined in the paragraph 1. Classically, the generated wave $A_P'(x',y')$ in the image plane is obtained by Fourier transformation of $A_P(x,y)$:

$$A_P'(x',y') = FT\left[A_P(x,y)\right] = \iint_{x,y} B_R(x,y) \exp\left[ i \frac{2\pi}{\lambda} \delta(x,y) - i\, 2\pi(ux + vy) \right] dx dy = \mathbf{C}(u,v) \tag{6}$$

with u = x'/λD, v = y'/λD, and **C**(u,v) is a complex function whose modulus (in arbitrary units) and phase (in radians) is proportional to the diffracted wave. Then the expression of the wave amplitude $A_{Fx}$'(x',y') transmitted by the linear filter of equation (2) placed in the image plane will be:

$$A_{Fx}'(x',y') = B_{u_0}(u)\ t_A(u)\ \mathbf{C}(u,v) = B_{u_0}(u)\ \frac{1+u/u_1}{2}\ \mathbf{C}(u,v) \tag{7}$$

where $u_0 = x_0'/\lambda D$, $u_1 = x_1'/\lambda D$ and $B_{u_0}(u)$ is the 'boxcar' function of half-width $u_0$, equal to 1 if $-u_0 \le u \le +u_0$ and to 0 everywhere else. This function corresponds to the useful area of the filter, while the multiplying factor $t_A(u)$ stands for its linear amplitude variation. The latter is displayed on the Figure 2, accompanied with a limited choice of alternative transmitting functions. It must be noticed that the case when $|u_1| < |u_0|$ (or correspondingly $|x_1'| < |x_0'|$) is perfectly conceivable from a mathematical point of view. However this would imply the practical realization of filters whose transmissions are either negative or higher than 100%. We shall then assume that the transmission must always stay comprised between 0 and 1 whatever is the $x_1$' parameter, as can be seen in Figure 2.

[Figure 2: Profiles of some typical varying density filters ($x_1$' = 0.2 mm)]

The expression of the complex amplitude distribution $A_{Fx}(x,y)$ in the exit pupil plane after spatial filtering is then equal to the inverse Fourier transform of $A_{Fx}$'(x',y'):

$$A_{Fx}(x,y) = FT^{-1}[A_{Fx}'(x',y')] = \iint_{u,v} B_{u_0}(u)\frac{1+u/u_1}{2}\mathbf{C}(u,v)\exp[i\ 2\pi(ux+vy)]\,du\,dv \tag{8}$$

that can be rewritten as:

$$A_{Fx}(x,y) = FT^{-1}\left[\frac{1+u/u_1}{2}\mathbf{C}(u,v)\right] \otimes\ 2u_0\ \mathrm{sinc}(2\pi u_0 x) \tag{9}$$

where the symbol ⊗ denotes a convolution product with the sine cardinal function, which is the well-known Fourier transform of the boxcar function. Then, from relations (5-6) and by using the derivation theorem stating that:

$$FT^{-1}[u\,\mathbf{C}(u,v)] = \frac{1}{2i\pi}\frac{\partial}{\partial x}\left\{FT^{-1}[\mathbf{C}(u,v)]\right\} \tag{10}$$

an analytical expression of $A_{Fx}(x,y)$ can finally be deduced:

$$A_{Fx}(x,y) = \frac{1}{2}\left[\left[\left\{B_R(x,y) + \frac{D}{x_1'} B_R(x,y)\frac{\partial\delta(x,y)}{\partial x} - i\frac{\lambda D}{2\pi x_1'}\frac{\partial B_R(x,y)}{\partial x}\right\}\exp\left[i\frac{2\pi}{\lambda}\delta(x,y)\right]\right]\right] \otimes \frac{2x'_0}{\lambda D}\operatorname{sinc}\left(\frac{2\pi x'_0 x}{\lambda D}\right) \quad (11)$$

and the observed intensity $I_{Fx}(x,y)$ in the pupil image is equal to the square modulus of $A_{Fx}(x,y)$. However, it already appears that the latter formula will not allow to retrieve the simple relationship (3) predicted by geometrical optics, unless three assumptions are made:

- Assumption n°1: $B_R(x,y)$ is defined as the amplitude transmission function of the pupil, uniformly equal to 1 inside its useful area, and to 0 elsewhere. But in reality only the inner region of the pupil (where we seek to estimate the wavefront errors) presents some interest here. When restricted to this area, $B_R(x,y)$ can simply be written as $B_R(x,y) = 1$.
- Assumption n°2: the partial derivative $\partial B_R(x,y)/\partial x$ of the transmission function is uniformly equal to zero in the whole pupil plane, at the exception of the ring surrounding and defining the useful pupil area, where it will be equal to $-\infty$ or $+\infty$ (these are indeed positive or negative Dirac functions as illustrated in the Figure 3). Eliminating this circular contour over a one-pixel width should allow us to consider that the partial derivative is equal to zero everywhere else within the computing area.
- Assumption n°3: Finally, we must neglect the effect of the convolution product with the sine cardinal function. This approximation is all the more valid as $x'_0$ tends toward infinity and the sine cardinal becomes similar to a Dirac distribution. Practically, it implies that the spatial filter has infinite dimensions.

[Figure 3: Partial derivative of the pupil transmission function along X-axis]

Hence when the three previous assumptions are verified, the relations (3-4) can easily be recovered from Fourier optics theory. Unfortunately the third hypothesis is not acceptable from a practical point of view: firstly, because an infinite value of $x'_0$ is neither realistic nor desirable (since its optimal figure depends on the pupil sampling). Secondly, because the convolution product of the sine cardinal function of width $\lambda D/x'_0$ with the Dirac functions from which $\partial B_R(x,y)/\partial x$ is composed (see Figure 3) will generate high spatial frequency oscillations spreading over the whole pupil area. This effect is illustrated in the Figure 4 and clearly differs from the Gibbs phenomenon that is familiar to Fast Fourier Transform users. Moreover, the relation (11) shows that this crossed convolution term is proportional to $1/x'_1$, meaning that the resultant errors are increasing with the slope of the varying density filter. Conversely, most of the quoted authors reasonably stated that the best results should be obtained with high filter slopes since the contrast of the pupil image will be improved. Therefore a compromise must be found between both adverse tendencies.

[Figure 4: An example of reconstructed slope along X-axis (case n°1, pure defocus)]

In summary, we have established that in theory the proposed method suffers from an intrinsic (or 'bias') measurement error. According to Fourier optics, the slopes along the X-axis cannot be estimated exactly when applying the basic relationship (4) predicted by geometrical optics. Obviously, this is also true for the slopes along the Y-axis and the final reconstructed wavefront as well. In other words, even when using a perfect experimental apparatus the retrieved WFE should never be equal to the one originally transmitted by the tested optical system. The here above assumption n°2 suggests that this bias error can be minimized when reducing the radius of the pupil by one sampling point, but a residual inaccuracy will remain, which is proportional to the slope of the gradient amplitude filter. Having brought to light this difficulty, we shall now estimate what is the inherent measurement error of the method, in order to compare it to the typical performance of other current types of WFS. Numerical simulations seem to be best appropriate for that purpose, and are described in the following section.

## 4 Numerical simulations

In order to assess the performance of the studied WFE measurement method, we developed an IDL computer program allowing various numerical simulations, based on the major steps listed below:

1) Firstly, the wavefront error $\delta(x,y)$ to be measured is imported from an external file. It is considered as the reference to be finally compared with the reconstructed wavefront in step n°10. Four different 'typical' reference cases were considered, as described here after
2) The partial derivatives of the reference WFE along the X and Y axes are then calculated by the subroutines incorporated within IDL. They will serve as reference for an eventual comparison with the slopes estimations obtained in step n°7
3) The complex amplitude distribution $A_P'(x',y')$ in the image plane is evaluated using the direct Fourier transform in equation (6)
4) $A_P'(x',y')$ is multiplied by the varying density filter following the relation (7). This step is performed for two different orientations of the filter, respectively along the X and Y axes
5) According to the relation (8), the complex amplitude distributions in the pupil image plane are computed by inverse Fourier transformations
6) Then the observed intensities $I_{Fx}(x,y)$ and $I_{Fy}(x,y)$ are obtained by multiplying the amplitude distributions by their complex conjugates
7) From the observed intensities and their calibration factor $I_N$, the WFE slopes are computed by applying the basic relation (4) along both X and Y axes
8) Then the pupil rim is eliminated from the slope maps over a one-pixel width, by means of a simple multiplication with a 'pupil mask' map that has been pre-computed and stored in an external file. The mask also takes into account an eventual central obturation of the pupil

9) The wavefront error is reconstructed by applying one of the algorithms described by Southwell [14]: here we selected the 'zonal reconstruction of type A' mentioned in his paper
10) Finally, the reconstructed wave is compared with the original WFE $\delta(x,y)$, and their bidimensional difference map is evaluated.

It must be highlighted that the accuracy of the phase retrieval procedure was systematically verified by means of the reference slope maps calculated at step n°2, so that it was checked that the intrinsic errors of this algorithm are negligible with respect to those of the studied measurement method. Four types of reference WFEs are considered here, as shown in Figure 5 and Figure 6:

- Case n°1: This is a 'pure defocus' error, meaning that the optical system to be tested can be considered as perfect (or diffraction-limited), but that its image plane – or the measurement device itself – is slightly shifted along the Z optical axis. In that case the WFE presents a spherical shape and its slopes should be linear along the X and Y axes
- Case n°2: These are low spatial frequency defects (for example astigmatism or coma aberrations), engendered either by polishing errors and mechanical deformations of the optical components, or by an imperfect alignment of the whole optical system
- Case n°3: Here the defects are of high spatial frequency and were measured on a real manufactured mirror. The errors are essentially due to imperfect polishing and deformations of the optical surface near the three mechanical attachment points
- Case n°4: These are random defects, representative of the optical path disturbances engendered by turbulent atmospheric layers during ground-based astronomical observations

Simulations were applied to an optical system of exit pupil diameter 2R = 500 mm and where the distance D to the image plane is 5 m, thus having an equivalent aperture number of 10. The wavelength $\lambda$ of the monochromatic light is always taken equal to 0.6328 µm. Other input parameters are summarized in the Table 1. The obtained results are illustrated by two series of image strips represented in the Figure 5 and Figure 6. Their purpose is essentially illustrative, whereas the obtained numerical values are compiled in the Table 1.

[Figure 5: Reference, measured and difference slope maps along X (upper row) and Y axes (middle row) obtained for a pure defocus. The corresponding WFEs are shown on the bottom row (grey-levels are scaled to PTV values)]

[Figure 6: Reference, measured and difference WFE maps in the case of low-frequency (upper row), high-frequency (middle row) and random defects (bottom row). Grey-levels are scaled to PTV values]

In the Table 1 are indicated, for each of the four considered cases, the PTV and RMS values of their reference, reconstructed and difference maps, as well as a global estimate of the error percentages, which is a raw indicator of the precision of the method. It must be pointed out that only optimal results are shown here, meaning that:

- The half-width $x_0$' of the boxcar function in the image plane is adjusted to the pupil sampling according to classical rules of Fourier transforms
- For each different WFE, we have determined the optimal ratio $x_1$'/$x_0$' minimizing the residual measurement errors
- And finally, the pupil rim is systematically eliminated from the computing area by means of the 'pupil mask' mentioned in the step n° 8 of the WFE retrieval procedure

[Table 1: Simulation results for different wavefront errors]

At a first glance, the numerical results presented in the Table 1 seem fairly good, indicating that the absolute precision obtained on the reconstructed WFEs is very high. For example in the case of a pure defocus of one-wavelength amplitude (case n°1), the intrinsic error of the method is estimated around λ/120 and λ/1700 in PTV and RMS sense respectively. Other figures remain quite satisfactory even in the case of much more complicated surface shapes such as the high spatial frequency defects (case n°3) that was deemed as the most unfavorable. In fact the estimated accuracy is always better than λ/10 PTV and λ/100 RMS while the error ratios stay within a few percents. It should then be concluded that the proposed method actually is competitive with other current wavefront measurement techniques, such as phase-shifting interferometry or Shack-Hartmann sensors. However a few restrictions should be highlighted:

- Firstly, the removal of the pupil rim strongly benefits to the inherent measurement error of the method: our simulations showed that the gain factor varies between three and ten, depending on the considered WFE. This means that the performance is sensibly decreased on the full pupil.
- Secondly, these errors are closely related to $x_1$' as predicted in the paragraph 3: small figures of $x_1$' (equivalent to high filter slopes) lead to a dramatic performance loss due to the diffraction of the filter, while moderate slopes also tend to decrease the measurement accuracy – although in a slower proportion. Table 1 also shows that the optimal figures of $x_1$' depend on the wavefront errors being measured themselves: as a rule of thumb, the best results are obtained when $x_1$' and $x_0$' are of the same order of magnitude.
- Finally, it must not be forgotten that these simulations are only applicable to a pinhole source of light, while real sources generally have a weak spatial area that should naturally smooth the measured slopes and reduce their high frequency components. However, this effect could either be favorable or detrimental to the method, depending on the WFE to be measured.

Having developed the detailed numerical model of a wavefront sensor based on a linear amplitude transmission filter naturally led us to consider other filter shapes, such as the Horwitz formula [9] or the never mentioned sine transmission functions (both in amplitude or intensity) represented on the Figure 2. In particular, the Horwitz and sine intensity profiles seem of particular interest, since their practical realization could be much easier (the linear intensity filter of Horwitz is a standard and not very expansive component. On their side, sine intensity profiles could be efficiently produced by means of holographic techniques). Such alternative filters can be modeled quite easily by simply adjusting the reference formulae (2) and (4) in our computer code. The results of this comparison are given in the Table 2 for a one-wavelength defocus (case n°1). They show that the measurement accuracy tends to decrease by a factor 2.5 when using the linear intensity filter, while it is not significantly modified by the sine transmission filters.

[Table 2: Simulation results for different filter shapes (case n°1)]

Finally, this set of numerical simulations allowed us to estimate the intrinsic measurement error of the method in a few typical cases of wavefront errors and filter shapes. The obtained results look very promising and show that the achievable accuracy is comparable with the best currently available technologies (laser-interferometers, Shack-Hartmann sensors), provided that certain precautions are taken (e.g. pupil rim removal or filter slope adjustment).

## 5 Conclusion

In this paper was discussed the theory and expected performance of a new class of wavefront sensors recently proposed by different authors. Basically derived from the Foucault test, their principle consists in evaluating the slopes of the wavefront emerging from a tested optical system by means of varying density filters placed into its image plane. A software tool reconstructing the WFE from its slopes completes the system. One decisive advantage of the method is the fact that light sources of different wavelengths or spectral widths can be employed. Also, sources of extended spatial area may be used within certain limits, as long as the resolution in the exit pupil of the optical system remains acceptable. Finally, different wavefront reconstruction algorithms compatible with real-time or a posteriori data-processing can be integrated into the wavefront sensor.

We showed that the proposed method can be envisaged following two different approaches. On one hand, elementary considerations of geometrical optics lead to a straightforward inversion formula between the intensity observed on the exit pupil and the wavefront slope, which has been adopted by most of the authors. On the other hand, Fourier optics theory provides a more complicated relationship, taking into account the diffraction effects generated by the filter slope and its limited size, and thus evidences strong distortions at the pupil rim. Therefore it appeared that the method suffers from an intrinsic measurement error, which was estimated through an extensive set of numerical simulations described in the section 4. From them a few important lessons were learned:

- Firstly, the removal of the pupil rim from the computing area is crucial to the performance of the method
- Secondly, the optimal filter slope $1/x'_1$ results from a careful compromise between the contrast of the observed intensities and the natural effects of diffraction
- When both previous conditions are fulfilled, simulations show that the intrinsic measurement error is well below $\lambda/10$ PTV and $\lambda/100$ RMS, which is comparable with the best currently available wavefront measuring techniques
- Finally, various filter shapes can be envisaged. The most promising might be the sine intensity filter, which combines the advantages of a good performance and an affordable realization process.

Hence the ultimate potential of this alternative class of wavefront sensors has been demonstrated, and clearly delimited. The next steps will logically consist in a thorough analysis of the instrumental errors (including in particular the spatial area and coherency of the light source, spatial non-uniformity of the incident beam, noises and non-linearity of the CCD detector, etc.) and the practical realization of a WFS prototype allowing the experimental assessment of its real performance.

## References

[1] L. Foucault, 'Mémoire sur la construction des télescopes en verre argenté', Annales de l'Observatoire de Paris vol. 5, p. 197-237 (1859)

[2] R. G. Wilson, 'Wavefront–error evaluation by mathematical analysis of experimental Foucault-test data', Applied Optics vol. 14, p. 2286-2297 (1975)

[3] E. M. Granger, 'Wavefront measurements from a knife edge test', in *Precision Surface Metrology*, J. C. Wyant ed., Proceedings of the SPIE vol. 429, p. 174-177 (1983)

[4] D. E. Vandenberg, W. D. Humbel, and A. Wertheimer, 'Quantitative evaluation of optical surfaces by means of an improved Foucault test approach', Optical Engineering vol. 32, p. 1951-1954 (1993)

[5] M. Born and E. Wolf, *Principles of optics* (London, Pergamon, 6th ed., 1980)

[6] R. Ragazzoni, 'Pupil plane wavefront sensing with an oscillating prism', Journal of Modern Optics vol. 43, p. 289-293 (1996)

[7] R. A. Sprague and B. J. Thompson, 'Quantitative Visualization of Large Variation Phase Objects', Applied Optics vol. 11, p. 1469-1479 (1972)

[8] R. Hoffman and L. Gross, 'Modulation Contrast Microscope', Applied Optics vol. 14, p. 1169-1176 (1975)

[9] B. A. Horwitz, 'Phase image differentiation with linear intensity output', Applied Optics vol. 17, p. 181-186 (1978)

[10] J. C. Bortz, 'Wave-front sensing by optical phase differentiation', J. Opt. Soc. Am. A vol. 1, p. 35-39 (1984)

[11] J. E. Oti, V. F. Canales and M. P. Cagigal, 'Analysis of the signal-to-noise ratio in the optical differentiation wavefront sensor', Optics Express, vol. 11, p. 2783-2790 (2003)

[12] T. Szoplik, V. Climent, E. Tajahuerce, J. Lancis, and M. Fernández-Alonso, 'Phase-change visualization in two-dimensional phase objects with a semiderivative real filter', Applied Optics vol. 37, p. 5472-5478 (1998)

[13] H. Furuhashi, K. Matsuda, C. P. Grover, 'Visualization of phase objects by use of a differentiation filter', Applied Optics vol. 42, p. 218-226 (2003)

[14] W. H. Southwell, 'Wave-front estimation from wave-front slope measurements', J. Opt. Soc. Am. vol. 70, p. 998-1006 (1980)

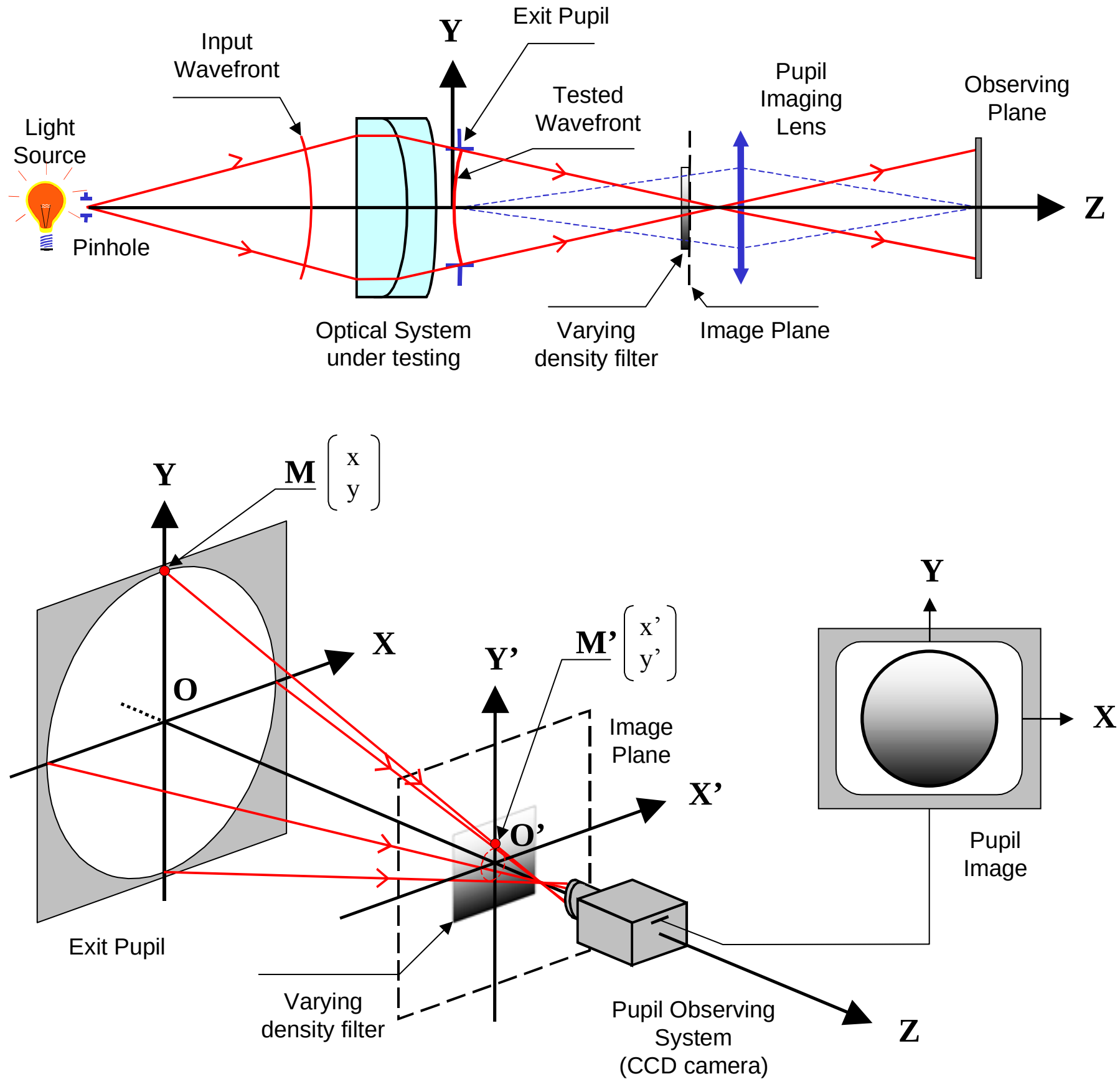

Figure 1: General principle of the wavefront sensor

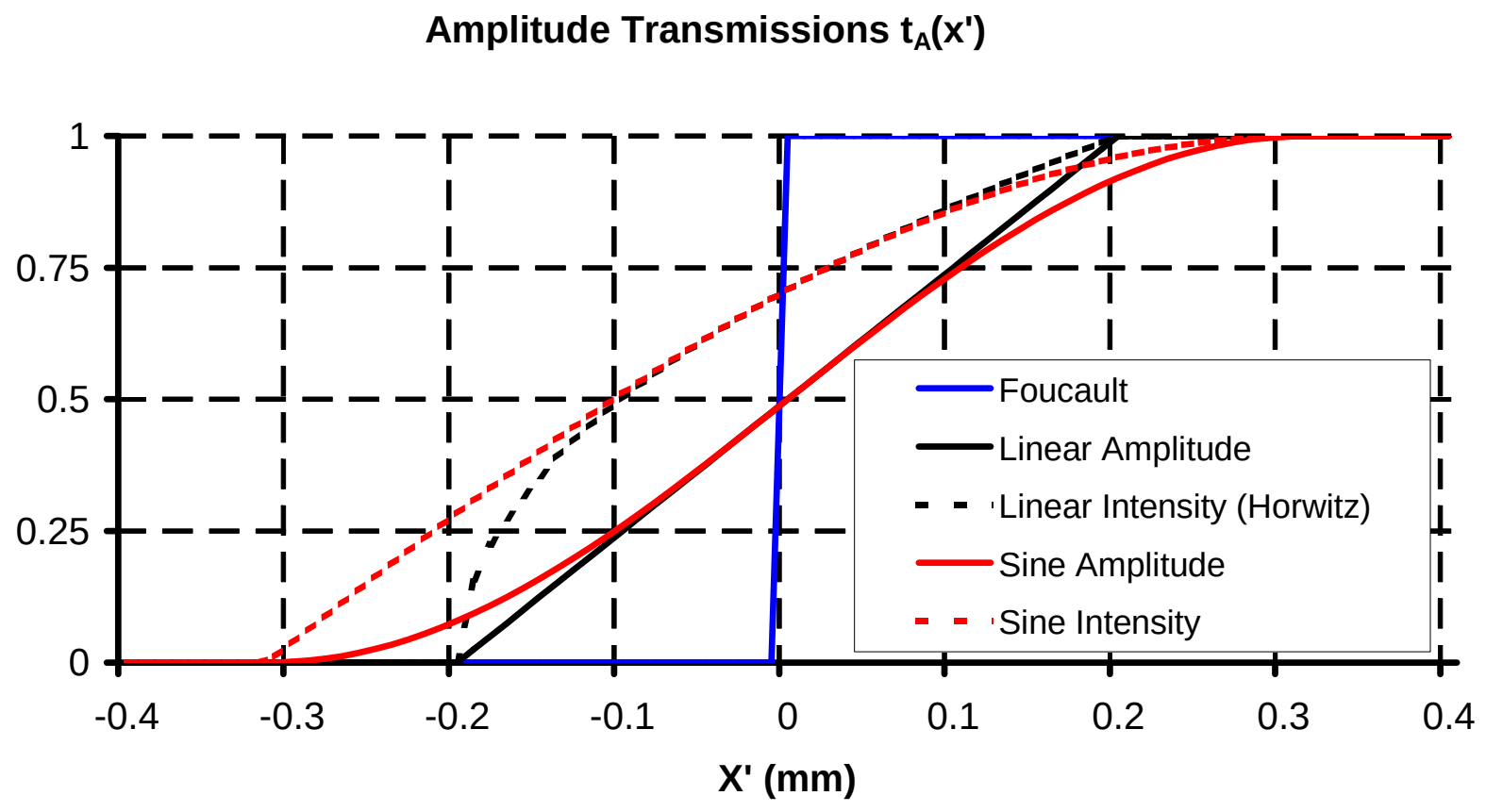

Figure 2: Profiles of some typical varying density filters ($x_1' = 0.2$ mm)

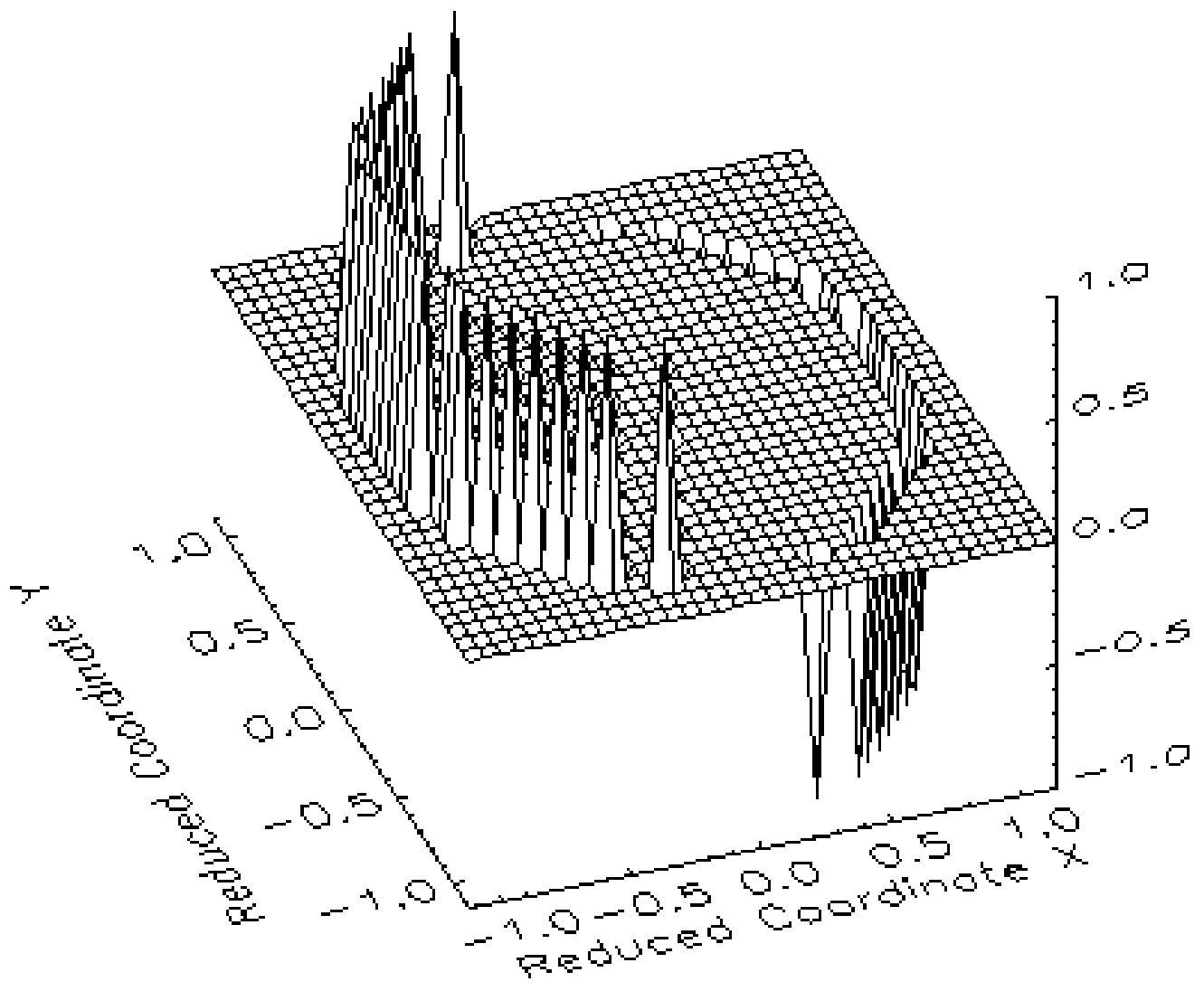

Figure 3: Partial derivative of the pupil transmission function along X-axis

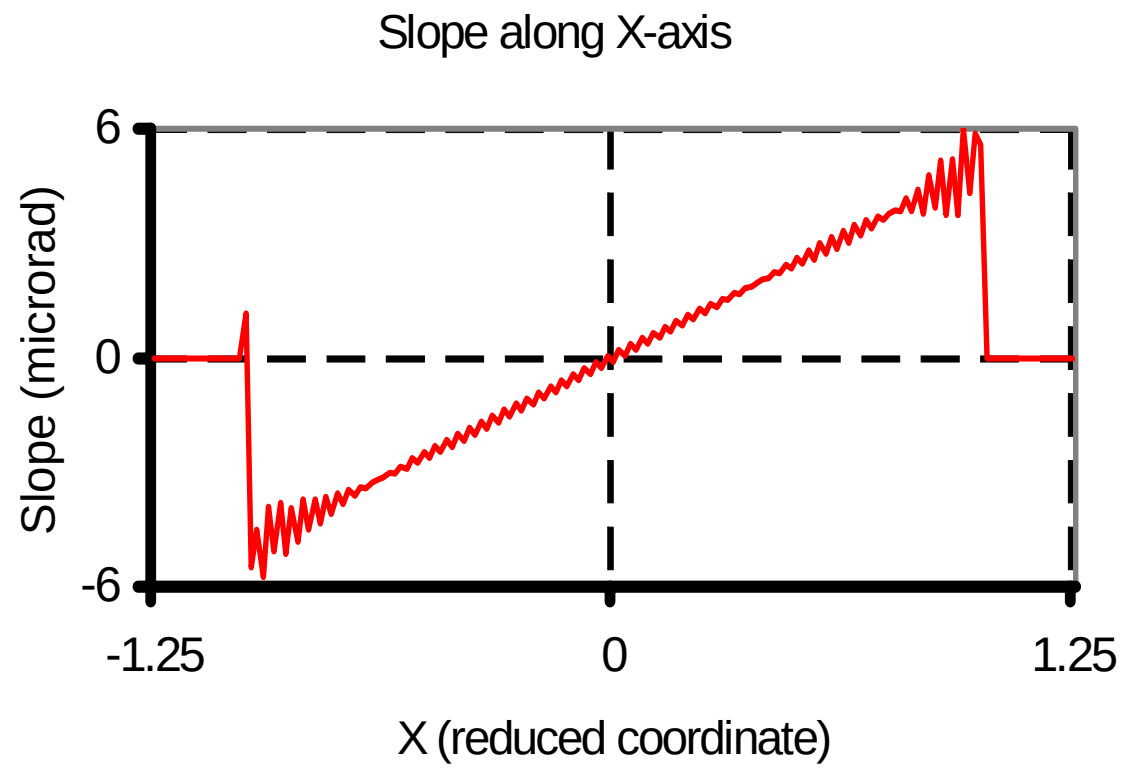

Figure 4: An example of reconstructed slope along X-axis (case n°1, pure defocus)

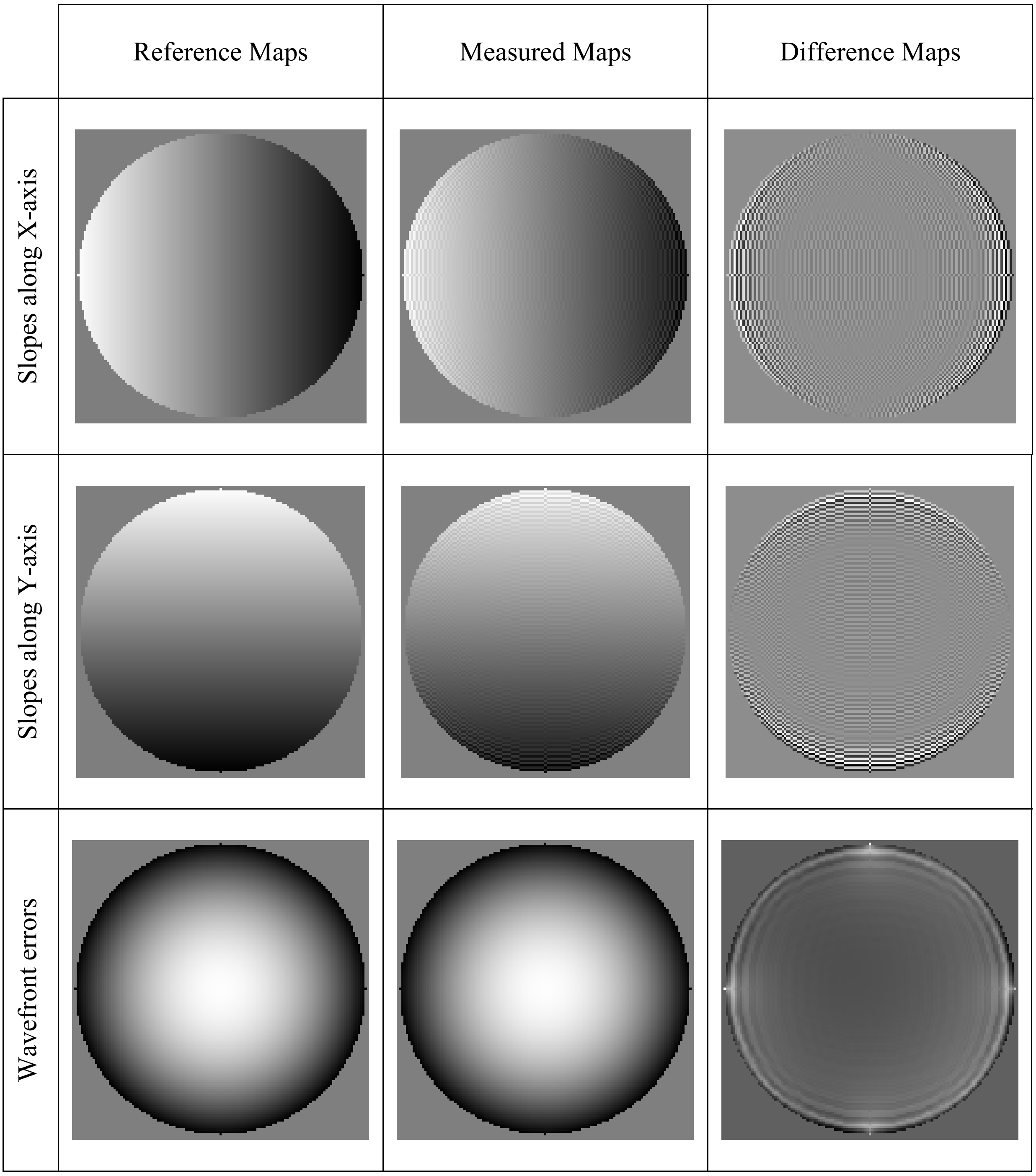

Figure 5: Reference, measured and difference slope maps along X (upper row) and Y axes (middle row) obtained for a pure defocus. The corresponding WFEs are shown on the bottom row (grey-levels are scaled to PTV values)

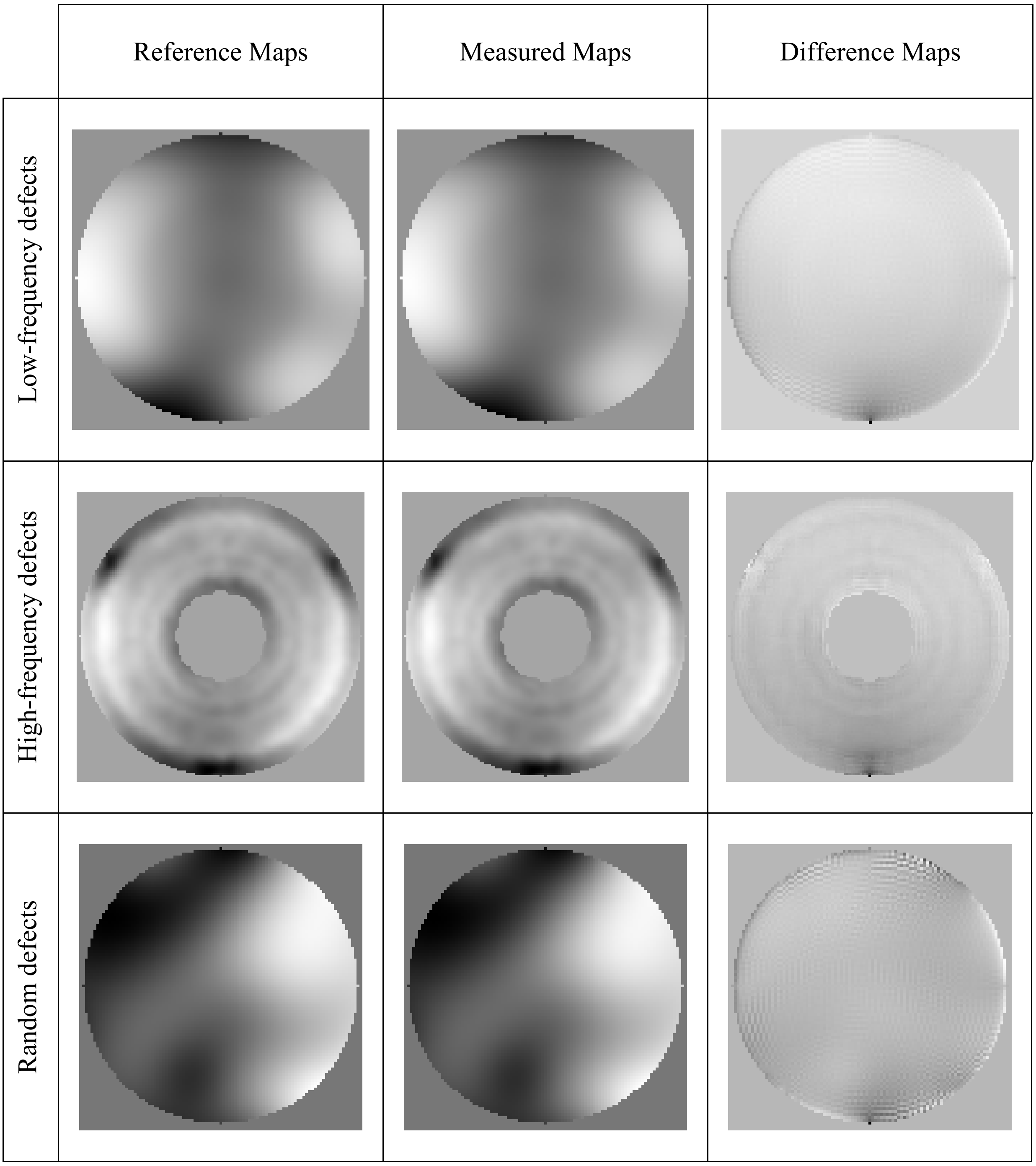

Figure 6: Reference, measured and difference WFE maps in the case of low-frequency (upper row), high-frequency (middle row) and random defects (bottom row). Grey-levels are scaled to PTV values

Table 1: Simulation results for different wavefront errors

| | | Case n°1: Pure defocus | Case n°2: Low spatial frequency defects | Case n°3: High spatial frequency defects | Case n°4: Random defects |
|---|---|---|---|---|---|
| INPUT PARAMETERS | Pupil sampling | 129 x 129 | 99 x 99 | 129 x 129 | 99 x 99 |
| | $x'_0$ (mm) | 0.4 | 0.31 | 0.4 | 0.31 |
| | $x'_1$ (mm) | 0.15 | 0.3 | 0.4 | 0.3 |
| REFERENCE MAPS | PTV (λ) | 0.956 | 1.675 | 1.258 | 3.351 |
| | RMS (λ) | 0.275 | 0.324 | 0.196 | 0.916 |
| MEASURED MAPS | PTV (λ) | 0.957 | 1.673 | 1.270 | 3.351 |
| | RMS (λ) | 0.275 | 0.324 | 0.197 | 0.916 |
| DIFFERENCE MAPS | PTV (λ) | 0.008 | 0.023 | 0.071 | 0.015 |
| | RMS (λ) | 0.001 | 0.002 | 0.005 | 0.001 |
| ERROR RATIO | PTV (%) | 0.9 | 1.4 | 5.6 | 0.5 |
| | RMS (%) | 0.2 | 0.5 | 2.7 | 0.1 |

Table 2: Simulation results for different filter shapes (case n°1)

| | | Linear Amplitude | Linear Intensity (Horwitz) | Sine Amplitude | Sine Intensity |
|---|---|---|---|---|---|
| REFERENCE MAPS | PTV (λ) | 0.956 | 0.956 | 0.956 | 0.956 |
| | RMS (λ) | 0.275 | 0.275 | 0.275 | 0.275 |
| MEASURED MAPS | PTV (λ) | 0.957 | 0.962 | 0.956 | 0.958 |
| | RMS (λ) | 0.275 | 0.275 | 0.275 | 0.275 |
| DIFFERENCE MAPS | PTV (λ) | 0.008 | 0.020 | 0.010 | 0.010 |
| | RMS (λ) | 0.001 | 0.002 | 0.001 | 0.001 |
| ERROR RATIO | PTV (%) | 0.9 | 2.1 | 1.0 | 1.0 |
| | RMS (%) | 0.2 | 0.6 | 0.2 | 0.2 |